%% file: iclr2019_conference.tex
\documentclass{article} 
\usepackage{iclr2019_conference,times}

\input{math_commands.tex}

\usepackage{hyperref}
\usepackage{graphicx}
\usepackage{url}
\usepackage{bbold}
\usepackage{caption}
\usepackage{subcaption}
\iclrfinalcopy 
\begin{document}\thispagestyle{empty}

\begin{center}

\LARGE\sc 
Ambulatory Atrial Fibrillation Monitoring \\ 
Using Wearable Photoplethysmography \\ 
with Deep Learning
\par
\end{center}



\author{Yichen Shen\thanks{ indicates equal contribution}, Alireza Aliamiri  \\
Samsung Strategy and Innovation Center\\
\texttt{\{yichen.shen,alireza.a\}@samsung.com} \\
\AND
Maxime Voisin\footnotemark[1], Anand Avati, Awni Hannun, Andrew Ng \\
Department of Computer Science \\
Stanford University \\
\texttt{maximev@stanford.edu} \\
}

%



\maketitle

\begin{abstract}
We develop an algorithm that accurately detects Atrial Fibrillation (AF) episodes from photoplethysmograms (PPG) recorded in ambulatory free-living conditions. We collect and annotate a dataset containing more than 4000 hours of PPG recorded from a wrist-worn device. Using a 50-layer convolutional neural network, we achieve a test AUC of 95\% and show robustness to motion artifacts inherent to PPG signals.
Continuous and accurate detection of AF from PPG has the potential to transform consumer wearable devices into clinically useful medical monitoring tools.

\end{abstract}

\section{Introduction}

Atrial fibrillation (AF) is the most common cardiac arrhythmia, affecting between 2.7 million and 6.1 million  adults in the United States. This number is expected to double over the next 25 years (\citeauthor{go2013heart}, \citeyear{go2013heart}).
AF is a risk factor for blood clots, cognitive impairment, heart failure and stroke (\citeauthor{kalantarian2013cognitive}, \citeyear{kalantarian2013cognitive}; \citeauthor{january20142014}, \citeyear{january20142014}).
The diagnosis is usually performed by observing the electrical activity of the heart in an electrocardiogram (ECG) typically measured with a cardiac event recorder, a Holter monitor or a chest patch.
However, these ECG devices tend to be used in a reactive manner rather than proactively. Many occurrences of
subclinical or silent 
AF are thus undetected, which results in a quarter of all ischemic strokes \citep{healey2012subclinical}.

Photoplethysmography (PPG) is an emerging technology that enables non-invasive heart rhythm measurement through optical sensing. A PPG sensor detects blood volume changes in the microvascular bed of tissue using a low intensity light. The optical mechanism PPG sensors use to measure blood volume change allows them to be placed in wearable devices like smartwatches. 

Using PPG sensors to detect AF has several advantages over ECG sensors. A PPG sensor can measure continuously and does not require active participation from the user, unlike ECG event recorders which must be activated by the user at the onset of any symptoms. Because of this, PPG can more accurately quantify AF burden, which is the percentage of time a subject's heart rhythm is in AF. AF burden is a much more indicative risk factor for heart attacks than the binary presence or absence of AF (\citeauthor{chen2018atrial}, \citeyear{chen2018atrial}; \citeauthor{go2018association}, \citeyear{go2018association}). 
Another advantage of PPG sensors is that they are already embedded in mainstream smartwatches which are deployed on mass scale. PPG-based monitoring via a smartwatch is seamless and can be activated over long periods of time with minimal discomfort compared to ECG-based monitors.
Hence, continuous AF monitoring using PPG sensors in mainstream smartwatches has the opportunity to be a more convenient, cost-effective solution to systematic, proactive AF screening. This would help detect challenging AF cases such as paroxysmal and silent AF which are often not diagnosed by opportunistic, reactive ECG-based AF screening \citep{freedman2017screening}.

PPG-based AF detection has received traction over the past few years.
Early attempts leveraged hand-crafted features about inter-beat intervals in PPGs (\citeauthor{tang2017identification, shan2016reliable, lee2012atrial, stankevivcius2016photoplethysmography, poh2018diagnostic, chan2016diagnostic, nemati2016monitoring}), while recent approaches trained deep neural networks on PPGs to detect AF (\citeauthor{shashikumar2017deep, aliamiri2018deep, gotlibovych2018end}). 
However, PPGs used in these studies were collected in controlled environments often inside a hospital, or were only a few minutes long.  A few attempts were made to detect AF from PPG in ambulatory free-living conditions for prolonged periods of time. These approaches either obtained moderate performance \citep{tison2018passive}, or deleted a significant portion of PPG segments -- e.g at least 33\% of PPGs in \citep{bonomi2016atrial}. This is largely due to the presence of noise and motion artifacts which corrupt the PPG. As a result, previous attempts have not been able to accurately identify AF episodes in PPG collected in an ambulatory free-living setting for a prolonged period of time.

In this work, we present the first model to continuously and accurately detect AF episodes in PPG
collected in an ambulatory free-living setting. The model achieves an AUC of 95\% on the test set. Furthermore we do not discard any PPG segment and show robustness to motion artifacts. 
To achieve these results, we train a 50-layer convolutional neural network to detect AF on more than 4000 hours of PPG signals
collected from 81 patients. Our work can be used for challenging downstream tasks like measuring AF burden in ambulatory conditions. We hope this work helps to pave the way toward transforming wearable devices into medical grade diagnostic tools. 

\begin{figure}
    \vspace{-10 pt}
    \centering
    \hspace*{-2.9cm} 
    \includegraphics[width=0.7\textwidth]{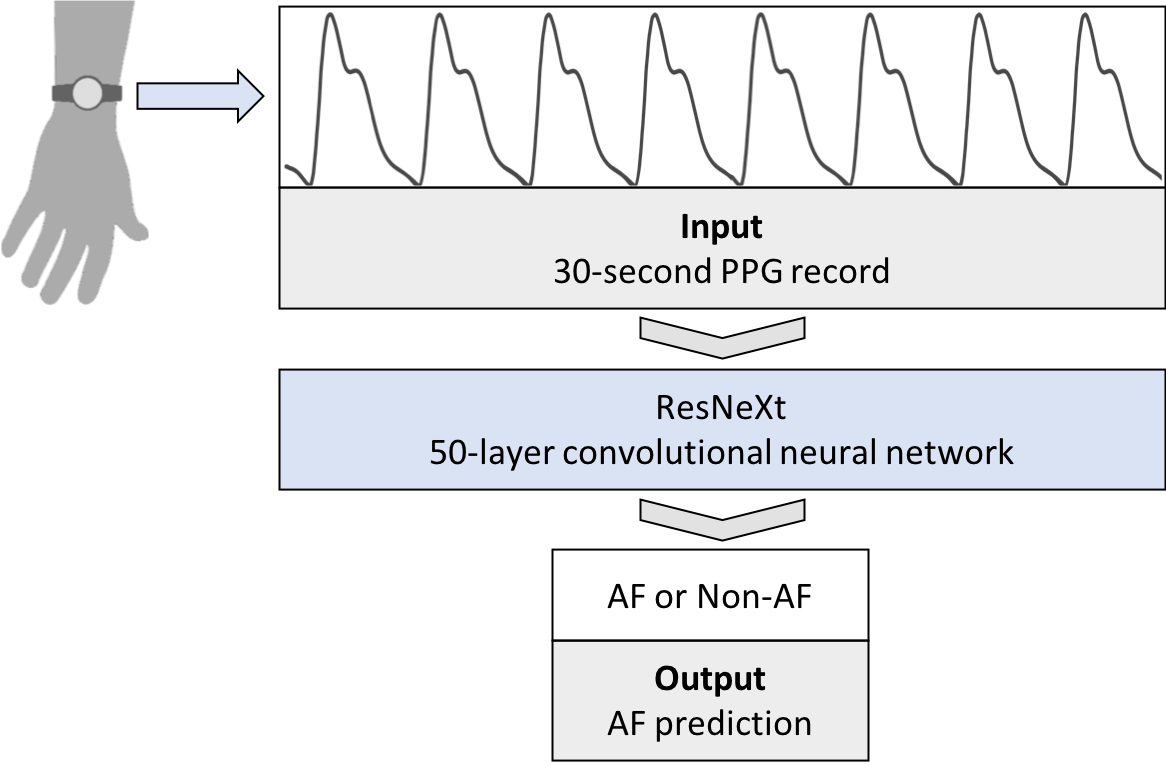}
    \caption{Our trained convolutional neural network correctly detects Atrial Fibrillation (AF) from other rhythms (Non-AF) on this PPG recorded with a wrist-wearable device}
    \label{fig:task}
\end{figure}

\section{Model}

\subsection{Problem Formulation}

Our goal is to detect AF episodes in a continuous PPG signal collected from free-living subjects. We extract consecutive 30-second records from the full PPG recording. For each 30-second record $x$, our model outputs a binary score $y \in \{0,1\}$ indicating respectively the absence or presence of AF. 
We optimize the binary cross-entropy objective function 

$$ \mathcal{L} = - \sum_{i=1}^N y^{(i)} \log p(y=1  \mid x^{(i)}) + (1-y^{(i)}) \log p(y=0  \mid x^{(i)}), $$

where $i$ is the index of the PPG record (there are $N$ records in total) and $p(y = l \mid x^{(i)})$ is the probability that the network
assigns to label $l$ given the input record $x^{(i)}$. 

\subsection{Model Architecture and Training} \label{model}

The AF prediction network is a 1D convolutional neural network (CNN). 
The input to the network is a 30-second PPG record sampled at 20 Hz. For each record we apply a Finite Impulse Response (FIR) low-pass filter with a cutoff frequency of 5Hz. We also scale the input record to have a mean of zero and a unit variance. The output of the network is a binary label indicating the absence or presence of AF in the input record.

\begin{figure}
    \vspace{-10 pt}
    \centering
    \includegraphics[width=1.0\textwidth]{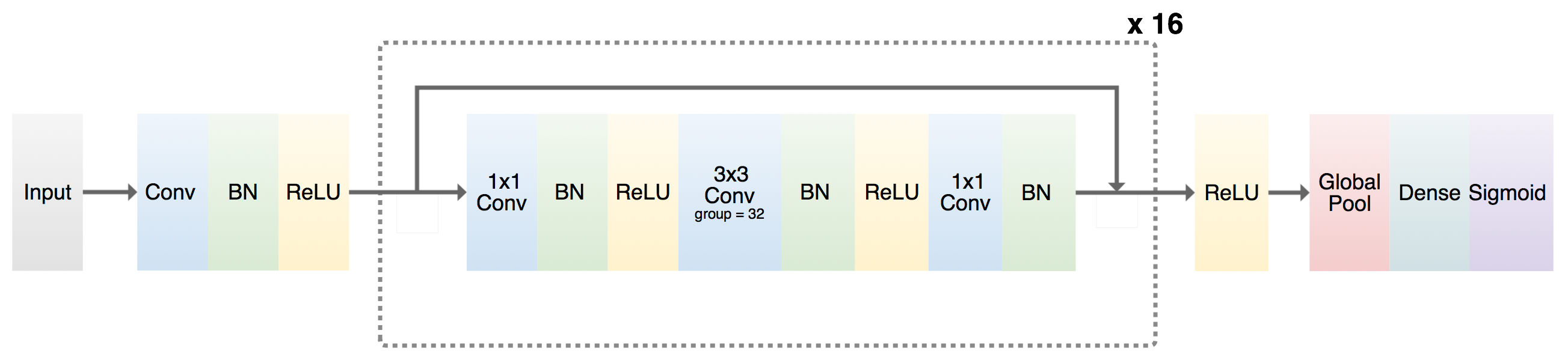}
    \vspace{-15 pt}
    \caption{The architecture of the network. The network consists of 49 layers of convolution followed by a global pooling layer, a fully-connected layer and a sigmoid.}
    \label{fig:architecture}
\end{figure}

The high-level architecture of the network is shown in Figure \ref{fig:architecture}. The CNN consists of 49 layers of 1D convolutions. This is followed by a global average pooling layer, a dense layer and a sigmoid layer to produce an output between 0 and 1.

The network consists of 16 ResNeXt bottleneck blocks (\citeauthor{xie2017aggregated}, \citeyear{xie2017aggregated}) sharing the same topology. 
The blocks rely on grouped convolutions which yield higher representation power than other state-of-the-art convolutional networks with the same number of parameters.
Each block consists of 3 convolutional layers. First, a 1x1 convolutional bottleneck layer reduces the number of feature maps. It is followed by a 3x3 grouped convolution which provides more expressive power in each block. Finally, a 1x1 convolutional layer restores the original number of feature maps.
These 16 blocks are grouped into 4 stages containing 3,4,6 and 3 blocks respectively. The spatial map is downsampled at the 3x3 grouped convolution layer of the first block of each stage.
The ResNeXt architecture was initially designed for 2D data. We adapt it for our 1D data. 
To ensure that each block has roughly the same computational complexity in terms of FLOPs, we downsample the spatial map using stride-4 convolutions -- rather than stride-2 convolutions in the 2D architecture -- at the 3x3 grouped convolution layer of the first block of each stage. We also remove the initial pooling layer to avoid downsampling the input record by too much.
Finally, we fine-tune the cardinality -- number of groups in the 3x3 grouped convolution -- and the bottleneck width of each block. Our best performing network has a cardinality of 32 and a bottleneck width of 4$d$, where $d$ starts out as 1 and is incremented at each stage of the network.

In order to make the optimization of such a network tractable, we employ shortcut connections in a similar manner to those found in ResNeXt architectures (\citeauthor{xie2017aggregated}, \citeyear{xie2017aggregated}). The shortcut connections between neural network layers optimize training by allowing information to propagate well in very deep neural networks.

Batch normalization (\citeauthor{ioffe2015batch}, \citeyear{ioffe2015batch}) and a rectified linear activation are applied after each convolutional layer.
We train the network from scratch, initializing the weights of the convolutional layers as in (\citeauthor{he2015delving}, \citeyear{he2015delving}). We use the Adam optimizer (\citeauthor{kingma2014adam}, \citeyear{kingma2014adam}) with the default parameters and minibatches of size 16. We save the best model as evaluated on the validation set during the optimization process.


\subsection{Baseline Model}
Inter-Beat Interval (IBI) is a feature commonly used for PPG-based AF detection. We implement the IBI algorithm described in (\citeauthor{lee2012atrial}, \citeyear{lee2012atrial}). For each 30-second PPG record, we identify beats in the PPG signal and compute the IBIs. The feature-based baseline algorithm then predicts AF by putting a threshold on the IBI variation measured in terms of Root Mean Square Successive Differences (RMSSD).

\section{Data} 
We used two datasets to train the model: the \textit{clinician-annotated} and the \textit{NSR} datasets. 
The \textit{clinician-annotated} dataset consists of 402 continuous PPG recordings collected from 29 free-living subjects. Each continuous PPG recording is 8 hours long on average. We simultaneously collected a reference ECG for rhythm annotation using an ECG patch. 
Out of the 29 subjects, 13 have persistent AF throughout their recordings, 2 have persistent normal sinus rhythm, and the remaining 14 display rhythms that change over time -- including 8 arrhythmias other than AF and normal sinus rhythm.
The \textit{NSR} dataset consists of 341 continuous PPG recordings collected from 53 healthy free-living subjects who self-reported as not having any symptoms of an arrhythmia. Each continuous PPG recording is 3 hours long on average. 
In summary, the two datasets in aggregate contain 743 continuous PPG recordings, each having a time span of a few hours.

All PPGs are recorded using a Samsung wrist-wearable device with a sampling frequency of 20 Hz. The device also records tri-axial acceleration, which is used in Section \ref{results_acceleration} to evaluate the model's robustness to motion artifacts.
In the clinician-annotated dataset, the reference ECG is collected from a single-lead, continuous monitoring patch with a sampling frequency of 500Hz. Each ECG is fully annotated by an ECG technician. The expert technician highlights segments of the continuous signal and marks them as corresponding to one of 10 rhythm classes: 8 heart arrhythmias, normal sinus rhythm and noise. All rhythms were labeled from their corresponding onset to offset, resulting in a full segmentation of the ECG.
The noise label is assigned when it is impossible to identify the underlying rhythm from the ECG. 

We break down the 743 continuous PPG recordings into 510,566 PPG records of 30 seconds. Each 30-second PPG record has one binary label which indicates if the 10-class rhythm segmentation of the corresponding ECG record contains AF. The binary label serves as the ground truth for training and evaluating models. PPG records whose corresponding ECG record is labeled as noise are discarded, since the ground truth rhythm is not known. They represent 1\% of the data.

The PPG records are split into a training, validation and test set. We ensure that there is no subject overlap between these sets. We also ensure that each set has almost balanced AF and non-AF records and that the proportion of subjects with respectively persistent AF, persistent normal sinus rhythm and multiple rhythms is similar across each set. The training, validation and test set contain respectively 42, 10 and 30 subjects in total, representing 50\%, 12\% and 38\% of the PPG records, as detailed in Table \ref{tab:sample-table}. 

The test set contains 147,968 records from the clinician-annotated dataset (10 test subjects) and 47,879 records from the NSR dataset (20 test subjects). Test subjects with persistent AF, persistent normal sinus rhythm and multiple rhythms represent respectively 45\%, 33\% and 22\% of the test records.  
50.2\% of the test records are labeled AF.

\begin{table}[t] 
\vspace{-10 pt}
\begin{center}
\begin{tabular}{c|ccccc}
   &\# train+val &\# train+val  &\# test &\# test \\
 Dataset &subjects &records &subjects &records\\\hline
 \rule{0pt}{2.6ex} Clinician Annotated        &19  &238345 &10 &147968 \rule[-1.ex]{0pt}{0pt} \\\hline
 \rule{0pt}{2.6ex} NSR   &32  &76374 &20 &47879\\
\end{tabular}
\end{center}
\caption{Two datasets were collected from free-living  ambulatory subjects. The PPG in the clinician-annotated dataset are fully annotated by clinicians using a reference ECG. The NSR dataset was collected from healthy subjects who report themselves as not having any arrhythmia. We give the total number of subjects and records for both the training and test sets. } \label{tab:sample-table}
\end{table}

\section{Results}

Models are compared based on their AUC, area under the Receiver Operating Characteristic (ROC) curve, which is independent of the prevalence of AF in the data. Each point on the ROC curve represents a sensitivity-specificity pair corresponding to a particular decision threshold. A model with high AUC enables practitioners to choose the sensitivity-specificity trade-off that best suits their use case. 
Models are compared based on AUC computed on the test set.

\subsection{Impact of motion artifacts} \label{motion-artifacts}

In ambulatory free-living conditions, motion artifacts are expected to corrupt the PPG and degrade the accuracy of AF predictions.
We evaluate the robustness of our model to such motion artifacts. To do so, a motion intensity score is assigned to each PPG record. This score is calculated as the standard deviation of the amplitude of the tri-axial acceleration.
PPG records in the test set are ordered by increasing motion intensity. 
We then evaluate the model AUC on the subset of test records whose motion intensity is in the lowest $c$-th percentile. By sweeping $c$ in $\{ 10,20,..100 \} $, an \textit{AUC-coverage curve} is created. Each point $(c, p)$ indicates that the model has an AUC of $p$ on the test records whose motion intensity is in the lowest $c$-th percentile.
Note that the AUC reported for a coverage $c=100\%$ is the AUC on the full test set.
A model robust to motion artifacts is expected to exhibit a flat AUC-coverage curve.

\begin{figure}
    \vspace{-10 pt}
    \centering
    \includegraphics[width=0.7\textwidth]{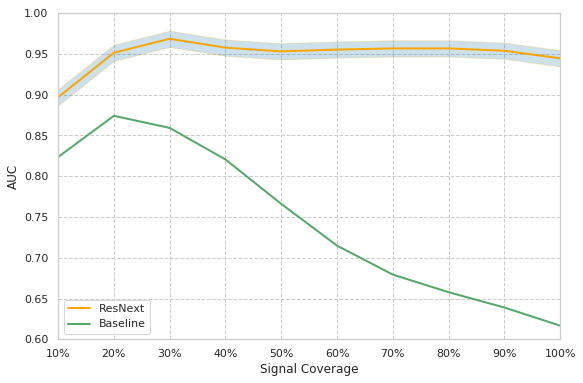}
    \caption{The AUC-coverage curves (see Section \ref{motion-artifacts}) of our deep learning model and of a baseline feature-based model. The AUC of the deep learning model does not degrade as predictions are done on test records with increasingly high motion intensity. This suggests that our deep learning model is robust to motion artifacts inherent to the ambulatory free-living setting. The performance of the deep learning model is averaged over 3 random seeds.}
    \label{fig:auccoverage}
\end{figure}

\subsection{Analysis} \label{results_acceleration}

\begin{figure}[!htb]
    \centering
    \includegraphics[width=0.45\textwidth]{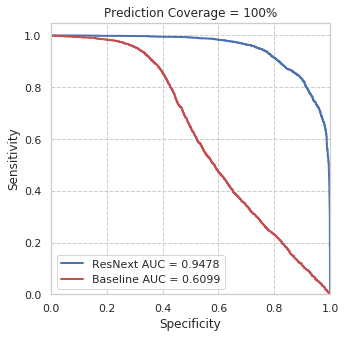}
    \caption{ROC curves (specificity vs sensitivity) of the deep learning model and a baseline model deployed on the full test set.}
    \label{fig:auc}
\end{figure}

The deep learning model obtains an AUC of 94.8\% on the test set.
The network largely outperforms the feature-based baseline. Interestingly, the performance of the deep learning model does not degrade when predicting on test records with higher motion intensity, unlike the baseline (see Figures \ref{fig:auccoverage} and \ref{fig:auc}). This suggests that our model is robust to motion artifacts typically encountered in free-living conditions.

Often the errors made by the deep learning model are understandable. First, although the model generally shows robustness against motion artifacts, it is still misled when too much noise corrupts the PPG. 
Second, we note in Figure \ref{fig:auccoverage} that performance decreases on PPG records whose motion intensity is in the bottom 20-$th$ percentile.
This is explained by the fact that low motion intensity does not necessarily correspond to clear PPG signal. 
For example, records collected from improperly worn wearable devices may have low motion intensity while not containing any relevant PPG morphology to predict AF. 
Finally, the AUC drops to 85.5\% on test patients with mixed AF and non-AF rhythms (partial AF). These records may be challenging to classify since they sometimes exhibit multiple arrhythmia other than AF as well as normal sinus rhythm. Also, only a handful of patients with partial AF are in the training set - so the model has relatively few such patients to learn from. These patients also
have noisier labels than persistent patients since the boundaries between different rhythms are fuzzy.

\section{Model Interpretation}

\subsection{Visualization of the learned low-level feature maps}

\begin{figure*}[t!]
    \vspace{-10 pt}
    \centering
    \begin{subfigure}[t]{0.9\textwidth}
        \centering
        \includegraphics[height=0.7in]{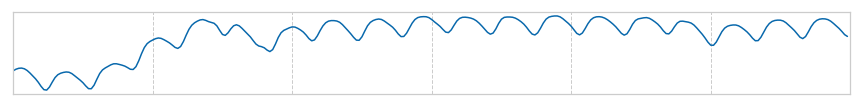}
        \vspace{-18 pt}
        \caption{15-second excerpt from a 30-second input PPG record}
    \end{subfigure}
    \hfill
    \begin{subfigure}[t]{0.9\textwidth}
        \centering
        \includegraphics[height=0.7in]{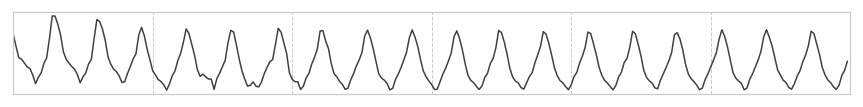}
        \vspace{-18 pt}
        \caption{Feature map learning the first order derivative of the input signal}
    \end{subfigure}
    \hfill
    \begin{subfigure}[t]{0.9\textwidth}
        \centering
        \includegraphics[height=0.7in]{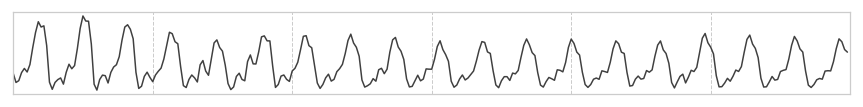}
        \vspace{-18 pt}
        \caption{Feature map learning the second order derivative of the input signal}
    \end{subfigure}
        \hfill
    \begin{subfigure}[t]{0.9\textwidth}
        \centering
        \includegraphics[height=0.7in]{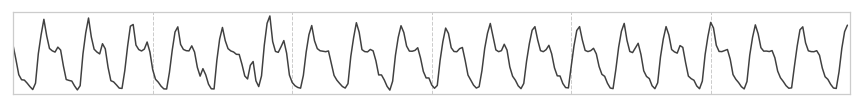}
        \vspace{-18 pt}
        \caption{Feature map identifying systolic and diastolic peaks}
    \end{subfigure}
    
    \vspace{-7 pt}
    \caption{Examples of feature maps learned by the model at the end of the first stage of the deep learning model, using (a) as input}
    \label{fig:featuremaps}
\end{figure*}

To understand which discriminative features the model uses to predict AF, we randomly select a test PPG record and visualize the feature maps obtained at the end of the first stage of the deep neural network. As shown in Figure \ref{fig:featuremaps}, the model learns to remove the low-frequency baseline wander. Some feature maps, shown in (d), identify systolic and diastolic peaks in the PPG. Having irregular beats is a symptom of AF. Other feature maps, shown in (b) and (c), seem to approximate the first and second order derivatives of the PPG signal. \citep{elgendi2012analysis} suggest that first and second order derivatives of a PPG signal contain information about the cardiovascular system such as hypertension and arterial stiffness, and \citep{cremer2015increased} suggests that the latter is an important predictor of AF in hypertensive patient. The features learned by our deep neural network therefore seem to be consistent with previous works in the medical field.

\subsection{Visualization of salient regions}

To further interpret the predictions of our deep learning model, we use the saliency mapping technique introduced by \citep{simonyan2013deep}. Saliency maps indicate which time steps influence most the prediction by computing the gradient of the binary cross-entropy loss with respect to each input time step. Formally, each time step $I_k$ in the input record $I = [I_0..I_N]$ has a saliency score $S_k$:
\[ S_k = |\frac{\partial L}{\partial I_k}|\]
where $L$ is the binary cross-entropy loss of the AF detection network for input record $I$.

\begin{figure*}[t!]
    \centering
    \begin{subfigure}[t]{1.0\textwidth}
        \centering
        \includegraphics[height=0.7in]{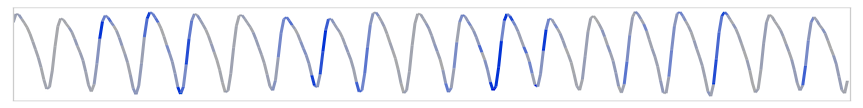}
        \vspace{-4 pt}
        \caption{Record with normal sinus rhythm. Predicted probability of AF = 0.0001}
    \end{subfigure}
    \hfill
    \begin{subfigure}[t]{1.0\textwidth}
        \centering
        \includegraphics[height=0.7in]{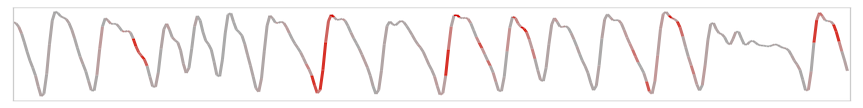}
        \vspace{-4 pt}
        \caption{Record with AF. Predicted probability of AF = 0.993}
    \end{subfigure}
    \vspace{-7 pt}
    \caption{Saliency maps of 15-second excerpts from two 30-second PPG records.
    Colored regions indicate salient regions which impact the model predictions. The model seems to focus on systolic and diastolic peaks as well as slopes to the left of the systolic peaks.
    }
    \label{fig:saliency}
\end{figure*}

Figure \ref{fig:saliency} provides the saliency mapping of two PPG samples. Colored regions are those with high saliency scores. They contribute most to the prediction. The network seems to focus on specific substructures in the PPG morphology such as systolic and diastolic peaks as well as slopes to the left of the systolic peaks. 

\subsection{Visualization of high-level feature space}
We visualize the high-level representations learned by the model from 8000 randomly selected test records, using the t-SNE method \citep{maaten2008visualizing}. The representations learned by the last convolutional layer of the network are mapped to a 2D space. The mapping is such that the joint probability of records close to each other in the high-dimensional representation space is similar to their joint probability in the 2D space.

\begin{figure*}[t!]
    \vspace{-10 pt}
    \begin{subfigure}[t]{1.0\textwidth}
        \centering
        \includegraphics[height=2.25in]{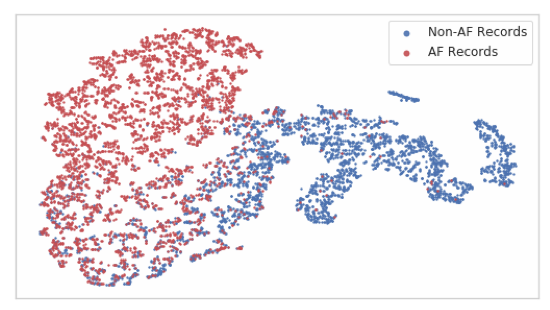}
        \vspace{-7 pt}
        \caption{ }
        \vspace{-5 pt}
    \end{subfigure}
    \begin{subfigure}[t]{1.0\textwidth}
        \centering
        \hspace{62 pt}
        \includegraphics[height=2.30in]{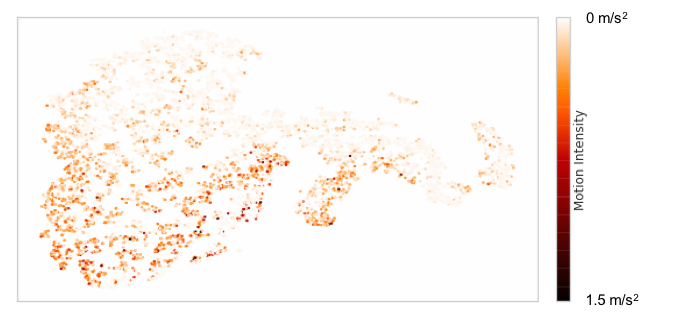}
        \vspace{-7 pt}
        \caption{}
        \vspace{-7 pt}
    \end{subfigure}
    \caption{(a) t-SNE plot with the label of each record and (b) t-SNE plot with the motion intensity of each record. AF and non-AF PPG records with low motion intensity are separable. PPG records with high motion intensity, mostly found in the bottom left quadrant, are more challenging to separate.}
    \label{fig:tnse}
\end{figure*}

Figure \ref{fig:tnse} provides the t-SNE visualization of the learned representations. 
In (a), we color records based on their ground-truth label. We observe that the cluster of AF records is mostly separable from the cluster of non-AF records.  In (b), we color the same records based on their motion intensity -- records with higher motion intensity have darker colors. We observe a continuous progression of motion intensity in the learned feature space. Records with high motion intensities are clustered in the bottom left quadrant of (b), whereas records with low motion intensities are in the top right quadrant of (b). By comparing (a) and (b), it appears that the cluster of PPG records with highest motion intensity is the hardest one to predict AF on.
This confirms that motion artifacts are a major challenge in PPG-based AF prediction in the ambulatory free-living setup.

\section*{Conclusion}

We develop a model which can accurately detect AF from continuous PPG records collected in the ambulatory free-living setting. Key to our approach is a large annotated dataset and a deep convolutional neural network. On the clinical side, future work should explore the possibility of predicting other types of arrhythmia and other forms of heart disease from PPG sensors available in wearable devices.
With the prevalence of inexpensive wearable devices, high-accuracy continuous arrhythmia monitoring from PPG can not only lower the risk of undiagnosed AF in general but also save precious time and resources from expert clinicians and cardiologists in resource intensive tasks like measuring AF burden. Furthermore, we hope that this technology can eventually provide accurate diagnostic information in places with constrained access to cardiologists and other medical resources.

\bigskip
\bigskip
\bigskip

\bibliographystyle{iclr2019_conference}
\bibliography{iclr2019_conference}

\pagebreak

\section{Appendix}

\begin{figure}[!htb]
    \centering
    \includegraphics[width=0.5\textwidth]{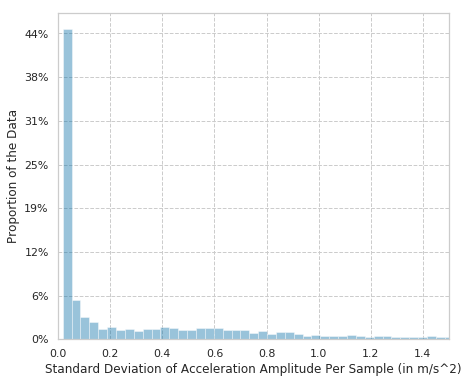}
    \caption{Distribution of motion intensity scores of the 30-second PPG records. Motion intensity is calculated as the standard deviation of the amplitude of the tri-axial acceleration.}
    \label{fig:accdist}
\end{figure}

\end{document}

%% file: math_commands.tex
\usepackage{amsmath,amsfonts,bm}

\def\eqref#1{equation~\ref{#1}}

\def\1{\bm{1}}

\DeclareMathAlphabet{\mathsfit}{\encodingdefault}{\sfdefault}{m}{sl}
\SetMathAlphabet{\mathsfit}{bold}{\encodingdefault}{\sfdefault}{bx}{n}

